\documentclass{emulateapj}
\usepackage{psfig}

\newcommand{\msun}{M$_{\odot}$}

\newcommand{\dlan}{DLA-B/FJ0812+32}

\begin{document}

\title{Constraints on Early Nucleosynthesis from the Abundance Pattern
of a Damped Ly$\alpha$ System at {\scriptsize \textit{z}} = 2.626}


\author{Yeshe Fenner}
\affil{Centre for Astrophysics \& Supercomputing,
Swinburne University, Hawthorn, Victoria, 3122, Australia}
\affil{UCO/Lick Observatory, University of California, Santa Cruz;
Santa Cruz, CA 95064}
\email{yfenner@astro.swin.edu.au}

\author{Jason X. Prochaska}
\affil{UCO/Lick Observatory, University of California, Santa Cruz;
Santa Cruz, CA 95064}
\email{xavier@ucolick.org}

\and

\author{Brad K. Gibson}
\affil{Centre for Astrophysics \& Supercomputing,
Swinburne University, Hawthorn, Victoria, 3122, Australia}
\email{bgibson@astro.swin.edu.au}

\begin{abstract}
  
We have investigated chemical evolution in the young universe by
analysing the detailed chemical enrichment pattern of a metal-rich
galaxy at high redshift. The recent detection of over 20 elements in
the gas-phase of a damped Lyman-$\alpha$ absorber (DLA) at $z = 2.626$
represents an exciting new avenue for exploring early
nucleosynthesis. Given a strict upper age of $\sim 2.5$~Gyr and a
gas-phase metallicity about one third solar, we have shown the DLA
abundance pattern to be consistent with the predictions of a chemical
evolution model in which the interstellar enrichment is dominated by
massive stars with a small contribution from Type~Ia supernovae.
Discrepancies between the empirical data and the models are used to
highlight outstanding issues in nucleosynthesis theory, including a
tendency for Type~II supernovae models to overestimate the magnitude
of the ``odd-even'' effect at subsolar metallicities. Our results
suggest a possible need for supplemental sources of magnesium and
zinc, beyond that provided by massive stars.

\end{abstract}


\section{Introduction}

Chemical abundances in local stars are fossil evidence from which
early nucleosynthetic processes may be inferred. An alternative probe
into the earliest epochs of the universe is via the direct detection
of metals in high redshift neutral gas. This requires a suitable
background source such as a quasar, whose spectrum reveals absorption
features in the intervening gas clouds. Damped Lyman-$\alpha$
Absorbers (DLAs) are one such class of quasar absorption system,
defined as having an H\,\small{I} column density
$N$(H\,\small{I})~$\gtrsim 2 \times 10^{20}$~cm$^{-2}$.

The recent discovery of an intervening galaxy along the sightline to
the quasar FJ081240.6+320808 (hereafter \dlan) has lead to the most
comprehensive abundance pattern measured beyond the local universe
(Prochaska, Howk, \& Wolfe 2003, hereafter PHW03). The gas-phase
abundances of over 20 elements - including B, Cu, Ga, and Ge - have
been detected in this galaxy, which has an unusually high column
density and metallicity.  Many of these elements have never before
been measured at high redshift and provide a unique opportunity to
probe early conditions of galaxy formation.

Identifying the modern-day counterparts to DLAs is still an open
question. Kinematic evidence is consistent with DLAs being large
protothick disks (Prochaska \& Wolfe 1997) or smaller merging
protogalactic clumps (Haehnelt et al. 1998), while chemical properties
have lead others to favour intervening dwarf or irregular galaxies
(e.g. Lanfranchi \& Fria{\c c}a 2003 and references therein). Galaxies
giving rise to DLAs are likely to encompass a wide size and
morphological range, from dwarf irregulars to giant
ellipticals. Indeed, imaging studies of low redshift DLAs expose a
variety of galactic types (e.g. Le~Brun et~al. 1997; Rao et~al.
2003). Observed DLAs may tend to probe different types of galaxies at
different evolutionary stages as a function of redshift.

The metal-line absorption profiles of the galaxy \dlan\ provide a
snapshot of the chemical composition in the interstellar gas along
only one sightline at one point in time.  However a sufficiently
detailed abundance pattern opens a window into the past history of
star formation because chemical elements are produced on different
characteristic timescales depending on their nucleosynthetic
origin. The abundance of metals in the gas-phase of galaxies and
protogalaxies reflects the interplay between processes including star
formation, the synthesis of elements by stars, the release of
metal-rich material via stellar winds and supernovae, and the flow of
gas. We have constructed a range of simple chemical evolution models
that follow these processes as a means of analysing this DLA. We
discuss instances where the inability to correctly predict specific
elemental abundances may indicate gaps in our understanding of
nucleosynthesis.

In Section~\ref{sect_obs} we describe the observations of the DLA at
$z$~=~2.626 along the sight-line to FJ081240.6+320808, upon which this
study is based. Section~\ref{sect_model} describes the ingredients of
the chemical evolution code used to model this DLA. In
Section~\ref{sect_results} the theoretical predictions are compared
with the DLA data and we discuss the sensitivity of the results to
age, metallicity, and model ingredients.

\section{Observations}\label{sect_obs}

\dlan\ was identified in the discovery spectra of the FIRST Bright
Quasar Survey (White et al.\ 2000). Subsequent $R \approx 8000$
spectroscopy with the ESI spectrometer (Sheinis et al.\ 2002) revealed
its very strong metal-line transitions (Prochaska et al.\ 2003).
PHW03 published HIRES echelle spectra of this `metal-rich' DLA and
reported the detection of over 20 elements.  The resulting abundance
pattern provides an unparalleled laboratory for the study of elemental
production in the young universe.

PHW03 derived gas-phase abundances using standard techniques and
estimated uncertainties in the measurements by propagating the
statistical errors.  With only a few exceptions, the dominant
uncertainty in the resulting elemental abundances is the effect of
differential depletion.  Similar to the ISM of the Milky Way,
refractory elements like Ni, Fe, Cr may be depleted from the gas-phase
such that their observed abundances are significantly lower than their
intrinsic values.  This is the principal challenge to studying
chemical abundances in the DLA and it has led to competing
interpretations of their chemical evolution history (e.g. Lu et al.\
1996; Kulkarni, Fall, \& Truran\ 1997; Prochaska \& Wolfe 2002;
Vladilo 2002).  In the following analysis, we adopt the conservative
dust corrections imposed by PHW03 based on empirical depletion
patterns observed for the Milky Way ISM (e.g.\ Savage \& Sembach
1996). Table~\ref{abund_table} lists both the gas-phase abundances and
dust-corrected abundance ratios for the elements observed in \dlan\ by
PHW03, along with statistical errors and dust-correction factors.
When relevant, we discuss the implications of dust depletion for our
conclusions. Further detail regarding the observational errors will be
given in Prochaska et al. (2003, in preparation).

\begin{table}
\caption{Elemental Abundances in \dlan\ $^{a}$ \label{abund_table}}

\begin{center}

\begin{tabular}{l|cccc}
\hline
\hline
 El & [X/H]$^{b}$ & $\sigma _{N} ^{c}$ & $\delta _{DC}$(90\%
 c.l.)$^{d}$ & [X/S]$^{e}$ \\
\hline
B  &      $-$0.57   &  0.085  &      0.1 (0.05)    &  +0.3    \\
N  &  $>$ $-$2.24   &  0.058  &      0.0 (0.1)     & $>$ $-$1.47     \\
O  &      $-$0.54   &  0.101  &      0.1 (0.05)    &  +0.33      \\
Mg &      $-$0.78   &  0.053  &      0.3 (0.1)     &  +0.29    \\
Al &  $>$ $-$2.00   &  0.054  &   $>$ 0.5          & $>$ $-$0.73    \\
Si &      $-$0.91   &  0.053  &      0.3 (0.1)     &  +0.16    \\
P  &  $<$ $-$1.06   &  0.000  &   $<$ 0.3          & $<$ +0.01    \\
S  &      $-$0.87   &  0.050  &      0.1 (0.05)    &  0.0    \\
Cl &      $-$1.55   &  0.000  &   $>$ 0.0          & $>$ $-$0.78    \\
Ti &      $-$1.87   &  0.112  &   $>$ 0.7          & $>$ $-$0.4    \\
Cr &      $-$1.61   &  0.032  &   $>$ 0.7          & $>$ $-$0.14   \\
Mn &  $<$ $-$1.85   &  0.000  &      0.7 (0.1)     & $<$ $-$0.38   \\
Fe &      $-$1.69   &  0.017  &   $>$ 0.7          & $>$ $-$0.22   \\
Co &  $<$ $-$1.48   &  0.000  &   $>$ 0.7          & $>$ $-$0.01   \\
Ni &      $-$1.73   &  0.007  &   $>$ 0.7          & $>$ $-$0.26   \\
Cu &  $<$ $-$1.11   &  0.000  &   $>$ 0.7          & $<$ +0.36    \\
Zn &      $-$0.91   &  0.022  &      0.2 (0.1)     &  +0.06   \\
Ga &  $<$ $-$1.45   &  0.000  &      0.7 (0.1)     & $<$ +0.02   \\
Ge &      $-$0.92   &  0.035  &      0.3 (0.1)     &  +0.15   \\
As &     $<$ 0.26   &  0.000  &      0.0           & $<$ +1.03     \\
Kr &  $<$ $-$0.44   &  0.000  &      0.0 (0.1)     & $<$ +0.33     \\
Sn &  $<$ $-$0.27   &  0.000  &      0.0 (0.1)     & $<$ +0.5    \\
Pb &  $<$ $-$0.10   &  0.000  &      0.0 (0.1)     & $<$ +0.67    \\
\hline
\end{tabular}
\newline
\end{center}
{\scriptsize
$^a$Measurements taken by Prochaska, Howk, \& Wolfe (2003).\\
$^b$Gas-phase abundance on a logarithmic scale relative to solar,
where $N$(HI) = 10$^{21.35}$ cm$^{-2}$. \\
$^c$Statistical error on gas-phase abundances. \\
$^d$Dust-corrections and uncertainties estimated from depletions
patterns observed in Galactic gas.  \\
$^e$Dust-corrected abundances on a logarithmic scale relative to S. 
}
\end{table}

\section{The Model}\label{sect_model}

We simulated the chemical enrichment history of a DLA using the
chemical evolution code described in Fenner \& Gibson (2003). Under
this formalism, primordial gas is allowed to collect and form stars
that synthesise new elements.  An exponentially decaying gas infall
rate was adopted and there were no outflows of gas driven by galactic
winds. In this way, the models described below resemble different
regions of disk galaxies.  The evolution of the gas phase abundance
pattern reflects the cumulative history of the dynamic processes of
star formation, stellar evolution, and nucleosynthesis. The classic
set of equations governing these processes (described in the seminal
work of Tinsley 1980), have been numerically solved by defining
$\sigma_i(t)$ as the mass surface density of species $i$ at time $t$,
and assuming that the rate of change of $\sigma_i(t)$ is given by:

\begin{eqnarray}
\displaystyle\frac{d}{dt} \sigma_{i} (t) & = & 
 \displaystyle\int^{m_{up}}_{m_{low}}\psi(t-\tau_{m})\,Y_i(m,Z(t-\tau_{m}))
 \,\frac{\displaystyle\phi(m)}{m} \,\; dm \nonumber \\
  & & + \displaystyle\frac{d}{dt} \sigma_i(t)_{infall} \nonumber \\
  & & -X_i(t)\,\psi(t),
\end{eqnarray}

\noindent
where the three terms on the right-hand side of equation~(1)
correspond to the stellar ejecta, gas infall, and star formation,
respectively.  $\psi$ is the SFR, $Y_i(m,Z(t-\tau_{m}))$ is the
stellar yield of $i$ (in mass units) from a star of mass $m$ and
metallicity $Z(t-\tau_{m})$, $\phi(m)$ is the initial mass function,
and $X_i$ is the mass fraction of element $i$. By definition, the sum
of $X_i$ over all $i$ is unity, and the total surface mass density is
identical to the integral over the infall rate. $m_{low}$ and $m_{up}$
are the lower and upper stellar mass limits, respectively, and
$\tau_{m}$ is the main-sequence lifetime of a star of mass $m$. In
practice, the first term is split into three equations that deal
separately with low-mass stars, Type~Ia supernova (SN) progenitors,
and massive stars.

We now describe each of the model ingredients in turn, distinguishing
the inputs that are parameterised using simple analytical
prescriptions from the nucleosynthesis inputs, which are derived
numerically from first principles.

\subsection{Analytical Prescriptions}

\emph{Initial Mass Function (IMF):} The initial mass function
determines the relative birth rate of stars as a function of mass.
Because different mass stars leave unique chemical signatures on the
ISM and operate on different characteristic timescales, the precise
form of the IMF is a key factor driving the evolution of abundance
ratios.  This model assumed the Kroupa, Tout, \& Gilmore (1993) three
component IMF. The Kroupa et al. function has fewer stars in the low
and high mass ends of the distribution than the single power-law
Salpeter (1955) function. In Section~\ref{sect_imf} we estimate the
sensitivity of the results to changes in the upper mass limit of the
IMF.

\emph{Star Formation Rate (SFR):} A simple analytical law for the SFR
was adopted. Akin to the Schmidt (1959) law, we varied the SFR in
proportion to the square of gas surface density. The efficiency of
star formation was adjusted in order to investigate different
timescales for metal enrichment.

\emph{Infall Rate:} The gas infall rate was assumed to decay  
exponentially on timescales ranging between 3 and 9 Gyr. For the same
final metallicity, a shorter infall timescale corresponds to a younger
object. The star formation efficiency was also assumed to be higher in
the models that became metal-enriched the fastest - this is motivated
by the expectation that a deeper potential well corresponds to a
faster rate of collapse and a higher efficiency of star formation. We
note that it is the combination of both infall rate and star formation
efficiency that is of foremost importance and these two effects
determine the time taken to reach a specific metallicity. In the
present study, the timescale for metal enrichment is more important
than either the infall timescale or star formation efficiency
considered in isolation. For this reason, we present results as a
function of age and base our discussion on ``age-sensitivity''.

\subsection{Stellar Yields and Lifetimes}\label{sect_yields}

\emph{Intermediate and low mass stars:} We incorporate the results of
van den Hoek \& Groenewegen (1997) whose metallicity-dependent yields
are specified over the mass range 0.9--8\,M$_{\odot}$. These stars are
integral in the synthesis of C and N, however observational
constraints for these two elements are lacking for \dlan.  Thus this
stellar population will be largely ignored in the following
discussions. Low-mass stars may be the principle production site for
Galactic Pb (Travaglio et al. 2001) as well as producing modest
quantities of As, Kr, and Sn via s-process nucleosynthesis. The
abundance of these s-process elements may prove to be sensitive
indicators of age in high redshift protogalaxies because they are
generated in 1$-$4~{\msun} stars with typical lifetimes between a few
hundred Myr and a few Gyr. However, remaining uncertainties in the
theory of the s-process, as well as observational complications
regarding s-element contamination by companions stars, may limit the
use of Pb as an age indicator.  An investigation into s-process
enrichment in DLAs will be the focus of a forthcoming paper.

\emph{Type~Ia supernovae (SNe~Ia):} A recalculation of the 1986 W7
model (Thielemann, Nomoto, \& Yokoi 1986) by Iwamoto et~al. (1999) was
adopted. About 0.75\,M$_{\odot}$ of iron is ejected per event. Our
model assumed that 3\% of binary systems involving intermediate and
low mass stars result in SNe~Ia. This fraction provides a good fit for
the solar neighbourhood (e.g.  Alib{\' e}s, Labay, \& Canal 2001;
Fenner \& Gibson 2003).

\emph{Massive stars:} Stars more massive than 8~$-$~10~{\msun} that
end their lives in violent supernova explosions are responsible for
most of the metals in the cosmos. We investigated chemical evolution
using two different sets of SN~II yields: 1) the Woosley \& Weaver
grid (1995, hereafter WW95) covering a mass range 11--40\,M$_{\odot}$;
and 2) a set of 36 Type~II SNe models covering the mass range
11--40\,M$_{\odot}$ at six metallicities (Z~=~0, 10$^{-6}$, 10$^{-4}$,
10$^{-3}$, $6\times10^{-3}$, $2\times10^{-2}$) recently compiled by
Limongi \& Chieffi (2003, hereafter FRANEC) using the latest version
of the FRANEC code described in Limongi \& Chieffi (2002).

For WW95, we took the lower energy ``A'' models for stars $\le 25$ and
the higher energy ``B'' models for heavier stars. Taking note of the
suggestion by Timmes, Woosley, \& Weaver (1995) that the WW95 mass
cuts may have penetrated too deeply within the iron core, we have
uniformly halved the iron yields from these models \footnote{We note
  that this violates the self-consistency of the stellar models, since
  a shift in mass cut that halves the Fe yield is expected to also
  modify the yield of elements like Co and Ni.}.  The FRANEC models all
have arbitrary mass cuts corresponding to 0.09--0.1\,M$_{\odot}$ of
$^{56}$Ni.

The nucleosynthesis models only extend to 40 and 35\,M$_{\odot}$ for
WW95 and FRANEC, respectively. Since stars as heavy as
100\,M$_{\odot}$ are permitted to form in our chemical evolution model
(albeit ten times less frequently than 40\,M$_{\odot}$ counterparts),
the predicted yields were linearly extrapolated for stars outside the
mass grid.  Section~\ref{sect_imf} discusses the sensitivity of the
results to the assumed upper mass limit on the formation of stars.

Unless otherwise stated, subsequent comparison between theoretical and
observed elemental abundances will refer to the models calculated
using WW95. In Section~\ref{sect_wwvsfr} we show that while WW95 and
FRANEC make very different predictions regarding the magnitude of the
``odd-even'' effect, for most of the elements between O and Ni, for
which DLA measurements exist, our conclusions drawn from the WW95
models also apply to the FRANEC models.

\emph{Stellar lifetimes:} We adopt metallicity-dependent main-sequence
(MS) lifetimes calculated by Schaller et~al. (1992). Stars pollute the
interstellar medium with metals over the course of their evolution via
stellar winds and planetary nebulae, however this model assumes that
all the mass loss takes place at the end of the MS phase. This
simplification is not expected to influence our predictions, since the
enrichment pattern is mostly controlled by massive stars whose metals
are returned to the ISM almost entirely during the supernova
explosion.

As a word of caution, this type of model assumes that stellar ejecta
is uniformly mixed into the gas phase and therefore predicts
\emph{mean} abundance trends. In reality, the products of SNe not only
mix inhomogeneously with the ambient gas, but may also be associated
with regions in different ionisation states. Despite these caveats, a
homogeneous chemical evolution model is a reasonable representation of
this DLA for several reasons: firstly, the measured enrichment pattern
of this $z\,=\,2.626$ object reflects the average in the gas phase
along a line-of-sight; secondly, in order to reach a metallicity of
$\sim$~1/3 solar, numerous generations of supernovae must have
exploded, which will tend to ``smooth out'' the spatial distribution
of the heavy elements; and, thirdly, similar to the majority of DLA
(Prochaska 2003) there are only modest variations in the relative
abundances of the DLA across the observed velocity profile.  These
small variations are likely associated to differences in differential
depletion implying a chemically homogeneous system.

\section{Results}\label{sect_results}

\subsection{Model versus Data}

\begin{figure*}
%\epsscale{1.24}
%\plotone{ChemComp_WWvsDLA_wrtS.eps}
\plotone{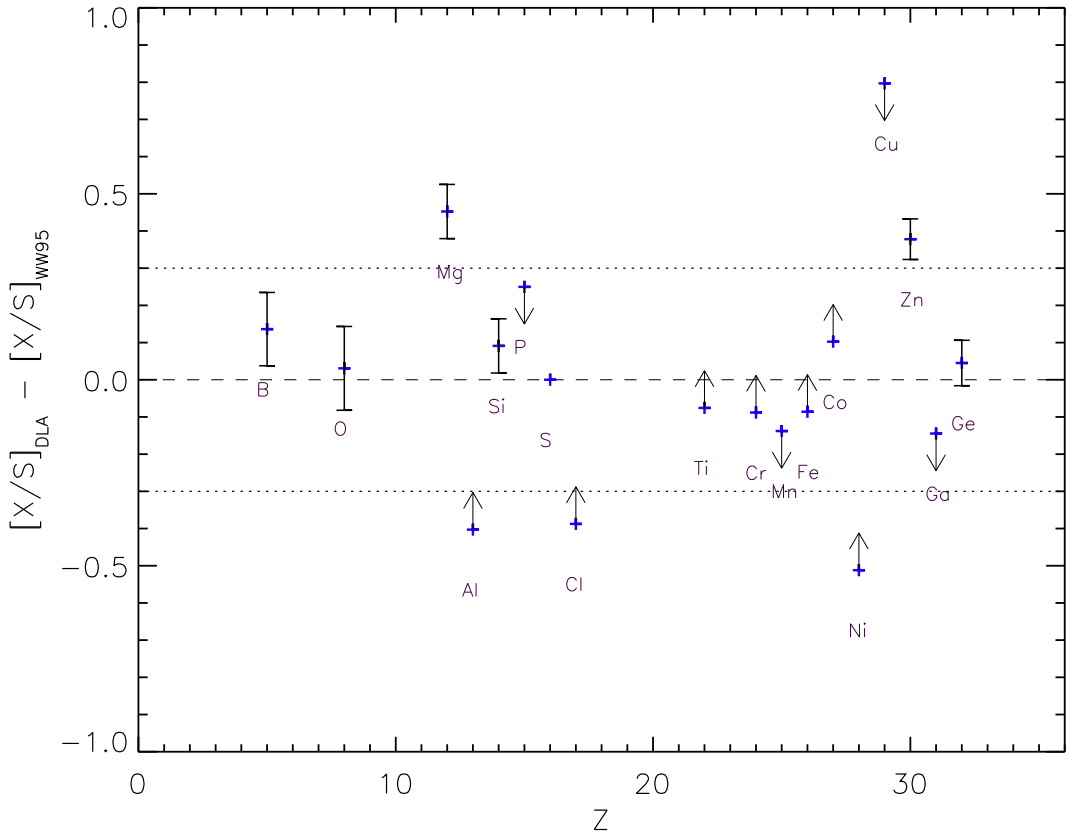}
\caption{Logarithmic ratio of the observed and predicted DLA abundance
  pattern using Woosley \& Weaver 1995 SN~II yields for a 2.2~Gyr
  model with [O/H]~$=$~$-$0.44. Arrows indicate the direction in which
  the difference will move if the upper and lower observational limits
  tighten. For elements like Fe, Ni, and Cr, the lower limits are set
  by dust depletion and are unlikely to change. Error bars reflect
  statistical errors on gas-phase abundance measurements. The level of
  uncertainty in the dust correction factor is comparable to the
  statistical errors. The dashed line indicates agreement between
  observations and predictions, while the dotted lines denote
  deviations of a factor of two.  \label{modelvsDLAfig}}
\end{figure*}

Figure~\ref{modelvsDLAfig} presents the difference between the
observed DLA enrichment pattern and the predicted [X/S]\footnote{We do
not use Fe as the reference element because its nucleosynthesis is
subject to many uncertainties relating to the SN~II mass cut and the
evolution of the SN~Ia rate.  Oxygen has a more robust theory of
production, but its observational status and solar abundance is less
certain. Thus we plot abundances relative to sulfur, for which an
ample database of DLA and stellar measurements exists. Moreover, the
nucleosynthetic origin of S is thought to be well understood - being
mainly generated by hydrostatic burning and explosive O and Si burning
in massive stars.} from our fiducial WW95 model that reaches
[O/H]~=~$-$0.44 after 2.2~Gyr. The dashed line indicates perfect
agreement between the model and the data, while the dotted lines
represent a factor of two difference.  We conservatively consider
predictions falling within the dotted lines to be acceptable.  The
arrows denote the direction in which the difference will move if the
observational limits tighten. For elements like Fe, Ni, and Cr, the
lower limits are set by dust depletion and are unlikely to
change. Error bars reflect statistical errors on gas-phase abundance
measurements, where the statistical error on [X/S] equals the
quadrature sum of the statistical errors on [X/H] and [S/H] given in
Table~\ref{abund_table}. When not reported as limits, uncertainties in
the dust corrections range from from 0.05 to 0.1~dex, i.e. comparable
to the magnitude of the statistical errors.

There is excellent agreement between the fiducial model and the DLA
data for most elements from B to Ge. This agreement is particularly
impressive given that the models we are applying to a high redshift
object were tuned using local stellar abundance patterns. Part of the
explanation is that the DLA pattern is close to solar.  Nevertheless,
the models match an observed trend in the $\alpha$-elements of
decreasing [X/S] with increasing atomic number (Figure~\ref{agefig}).
Furthermore, the model reproduces the mildly enhanced ``odd-even''
effect indicated by the limits to the Al, P, and Cl abundances.  There
is also good agreement for the iron-peak elements Cr, Mn, Fe and Co
and the heavier r-process elements Ga and Ge. The Ni and Cu
predictions do not violate the observational limits. The most serious
disagreement occurs for Mg and Zn, which are underpredicted by the
model by about 0.45 and 0.4~dex, respectively. Detailed discussions of
these elements are found in Sections~\ref{mg_sect} and \ref{zn_sect}.

\subsection{Model Sensitivity}

\subsubsection{Dependence on Age}

\begin{figure*}
%\plotone{ChemComp_wrtS_forDLAZ_ghz_jl.eps}
%\epsscale{1.25}
\plotone{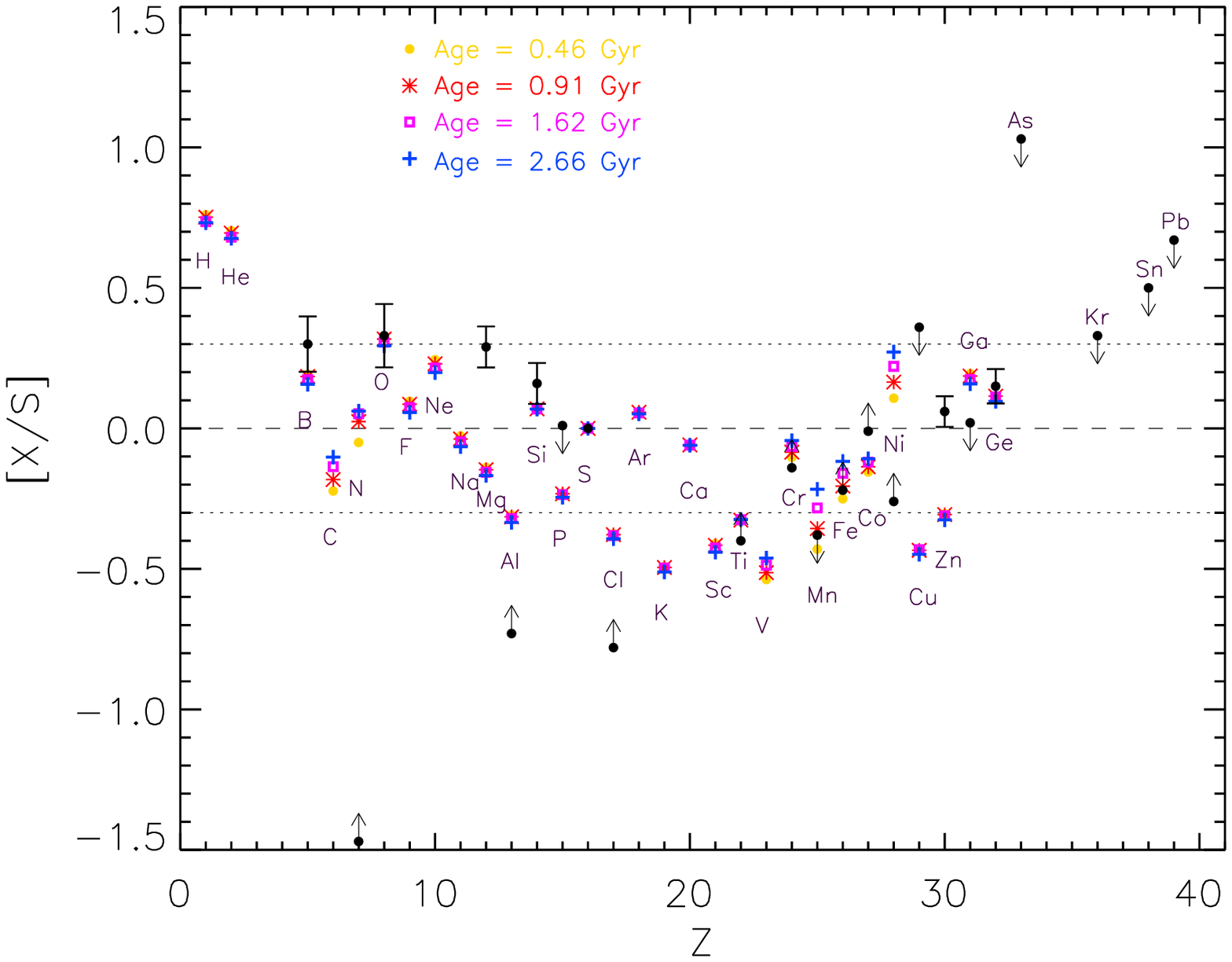}
\caption{Predicted abundance pattern  as a function
  of age. [O/H]~=~$-$0.44 in each model. Abundances are expressed
  relative to solar and scaled to sulfur, such that
  [X/S]~=~log$_{10}$(m(X)/m(S))$-$log$_{10}$(m(X)/m(S))$_{\odot}$.
  Four ages ranging from 0.46~-~2.66~Gyr are represented by the
  symbols indicated. The dust corrected DLA abundance pattern is
  indicated by black circles. Statistical errors on the gas-phase
  abundance are indicated by error bars. Arrows reflect the
  corresponding lower and upper limits. The dotted lines indicate
  deviations from scaled solar by a factor of two. \label{agefig}}
\end{figure*}

Figure~\ref{agefig} presents the predicted abundance pattern for a
model galaxy reaching a final metallicity of [O/H]~=~$-0.44$ after
four different periods: 0.46, 0.9, 1.6, and 2.66 Gyr. The
dust-corrected DLA abundance pattern (as presented in column 5 of
Table~\ref{abund_table}) is indicated by black circles, with arrows
specifying the upper and lower limits, where applicable.  Upper limits
on As, Kr, Sn, and Pb are shown, although their evolution is not
modelled. Abundances are on a solar logarithmic scale relative to
sulfur. The dashed line corresponds to the solar pattern and the
dotted lines are deviations by a factor of two. The detection of and
limits placed on 23 elements from B to Pb reveal a roughly solar-like
enrichment pattern.  This may not come as a surprise given the
moderate metallicity of the object.  However the redshift of this
galaxy imposes an upper age limit of $\sim$~2.5~Gyr, implying a much
shorter timescale for metal-enrichment than in the solar
neighbourhood.

The abundance of most elements relative to sulfur is virtually
constant over the age range shown in Figure~\ref{agefig}. This is
because the enrichment pattern is mostly set by short-lived massive
stars. The exceptions are carbon, nitrogen, and the iron-peak elements
where modest differences are predicted.  Low and intermediate mass
stars have long been held to dominate the production of N and possibly
that of C, although theoretical (e.g. Carigi 2000) and observational
(e.g. Henry, Edmunds, \& K{\" o}ppen 2000) studies also support a
strong contribution to C abundance from Type~II SNe.  Considerable
synthesis of C, by the triple-alpha reaction of helium and N, via CNO
processing, takes place within low and intermediate mass stars, which
are longer lived than SN~II progenitors.  Hence the enrichment of the
interstellar medium with C and N is delayed with respect to products
of massive star nucleosynthesis. We stress that the results presented
in this paper are largely insensitive to age.

The main deviation from the solar pattern is the mild enhancement of
the alpha-elements O, Mg and Si with respect to Fe or Zn. There is
also a trend toward lower relative abundances of $\alpha$-elements
with higher atomic numbers. Massive, short-lived stars culminating in
Type~II SNe are the primary source of the alpha-elements, while Fe is
produced in significant amounts by Type~Ia SNe, whose lower mass
progenitors have longer main-sequence lifetimes. Similar
alpha-enhancement is also seen in halo and thick disk stars where it
is understood to signify rapid star formation, whereby most stars
formed from gas that had not yet been polluted with iron-rich SN~Ia
ejecta. Indeed, the ratio of iron-peak elements to sulfur is 30-60\%
higher in the 2.66~Gyr model compared with the 0.46~Gyr model.

It is apparent from Figure~\ref{agefig} that a protogalaxy that has
reached 1/3 solar metallicity on a timescale of order 1~Gyr has
already been contaminated by the nucleosynthetic products of
intermediate mass stars and Type~Ia SNe. This is because intermediate
mass stars and SN~Ia progenitors can have lifetimes as short as a few
hundred Myr. Although the relative abundances of C, N, and the
iron-peak elements exhibit moderate age sensitivity in
Figure~\ref{agefig}, a better probe of age in an intermediate redshift
protogalaxy might be s-process elements such as Kr and Pb, since their
origin in low mass stars should ensure an enrichment timescale in
excess of $\sim$~1~Gyr (Travaglio et al. 2001). A preliminary
examination of Pb evolution indicates that [Pb/S] may increase by
roughly an order of magnitude between 0.5~and~2.5~Gyr (Fenner et al.
2004, in preparation). The anticipated detection of s-process elements
in additional DLAs may prove fruitful in constraining age.

\subsubsection{The Role of Type~Ia SNe}

Figure~\ref{agefig} reveals how the relative contribution from Type~Ia
SNe increases with the age of the object, for the same metallicity.
Type Ia SNe are believed to be the chief source of the iron-peak
elements. While the precise identity of SNe~Ia progenitors is debated,
they must involve intermediate to low mass stars in binary systems.
The explosion occurs when a white dwarf in the binary system accretes
enough material from its companion to exceed the Chandrasekhar
limit. The timescale for the SN~Ia event is thus determined by the
evolution of the companion star. The characteristic timescale for
SNe~Ia is often taken to be 1~Gyr, however this is
environment-dependent and moreover, the first Type~Ia events may occur
within only a few hundred Myr (Matteucci \& Recchi 2001).

In theory, the iron-peak abundances might be used to constrain the age
of this DLA. This requires assumptions to be made about the incidence
of SNe~Ia. As mentioned in Section~\ref{sect_yields}, we have adopted
the Milky Way value for the fraction of low and intermediate-mass
binary systems that culminate in SNe~Ia. In practice, there is a
conflict between the predictions for Mn, which imply an age $\lesssim$
1~Gyr and predictions for Fe, which imply that the protogalaxy has
been forming stars for more than 1~Gyr. In other iron-peak elements,
[Co/S] falls slightly below the lower limit in all models, while Cr
and Ni predictions agree with the data.

From Figure~\ref{agefig}, Mn appears to have a very strong dependence
on the age of the simulated galaxy. According to the nucleosynthesis
prescriptions adopted in this paper, SNe~Ia play such a strong role in
Mn production that their contribution exceeds that from SNe~II after
$\sim$~2.5~Gyr. It has been stated that Timmes et~al. (1995) claim
Type~Ia~SNe to be unimportant contributors to Mn synthesis (e.g.
Nissen et~al. 2000), yet it is clear from Figures 4 and 5 from Timmes
et~al. (1995) that Type~Ia~SNe produce $\sim$50\% of the solar Mn
abundance. What Timmes et~al.  showed instead was that the same trend
of [Mn/Fe] vs [Fe/H] can be obtained either with or without SNe~Ia due
to the strong metallicity-dependence of the WW95 SN~II Mn yields.
However, SNe~Ia are needed to reach the solar abundance. When the
same nucleosynthesis prescriptions used in this paper are applied to
the solar neighbourhood, we find that $\sim$75\% of the solar Mn
abundance originates from SNe~Ia.

The tendency for Mn to be overpredicted in these DLA models may
suggest that the main Mn source is metallicity-dependent. Other DLA
systems have been observed for which [Si/Fe]~$\sim 0$ yet [Mn/Fe] is
significantly subsolar (e.g. Pettini et~al. 2000).  Current
nucleosynthesis models cannot explain such behaviour, since a solar
Si/Fe ratio points to a Type~Ia SN contribution, but this is not
supported by the low Mn abundance, whose origin is understood to lie
mostly in SNe~Ia.  A similar trend has been identified in stellar
abundances of the Sagittarius dSph galaxy (McWilliam, Rich, \&
Smecker-Hane 2003).  Invoking metallicity-dependent SNe~Ia Mn yields
might help explain these apparently incompatible observations.

In a handful of DLAs with low dust content, Pettini et~al. (2000)
discern no correlation between [Mn/Fe] and [Fe/H], even though a clear
trend is apparent in local stars at all metallicities (Carretta et~al.
2002). They attribute this to the varied histories and morphologies of
the objects and suggest that a process in addition to Type II and Ia
nucleosynthesis may operate to explain Mn enrichment.

It should be noted that the theoretical yields of Mn from WW95 do not
have sufficient metallicity-dependence to account for the low [Mn/Fe]
ratios observed in very metal-poor stars (e.g. Alib{\' e}s, Labay, \&
Canal 2001). Furthermore, Co/Fe ratios up to four times higher than
solar are detected in local stars below [Fe/H]~$\sim -2.5$ (Cayrel
et~al. 2003), in conflict with the expectations of SNe~II
models. Thus, it is likely that gaps remain in our understanding of
the processes responsible for Mn and Co that limit the power of this
model to constrain the DLA age using these elements.

\subsubsection{Sensitivity to Metallicity}

\begin{figure*}
%\plotone{ChemComp_wrtS_ghz_jp.eps}
%\epsscale{1.25}
\plotone{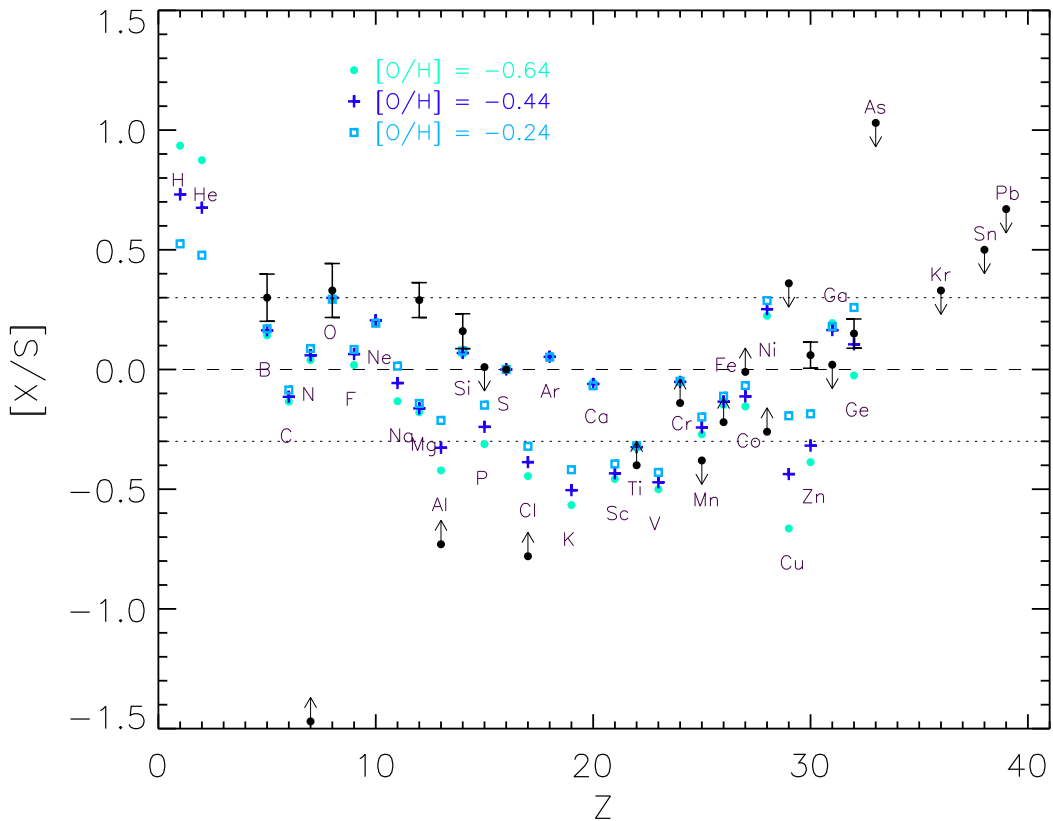}
\caption{Predicted abundance pattern as a function
  of metallicity. The enrichment pattern for 2.2~Gyr model galaxies at
  [O/H]~=~$-$0.44~$\pm$~0.2 are shown by the symbols indicated. Other
  symbols have the same meaning as in Figure~\ref{agefig}.
  \label{Zfig}}
\end{figure*}

The production of certain elements (particularly those with odd atomic
numbers) by massive stars is a strong function of the initial stellar
chemical composition. We examined the dependence of the predicted
enrichment pattern on the model galaxy's metallicity.
Figure~\ref{Zfig} depicts results from three models reaching
[O/H]~=~$-0.24, -0.44$, and $-0.64$ after 2.2~Gyr of star formation.
The $\pm$~0.2~dex variation in [O/H] reflects the uncertainty in the
measured H\,\textsc{i} column density.

\begin{figure*}
%\plotone{ProdFactors_WW1995_100Mup.eps}
%\epsscale{1.28}
\plotone{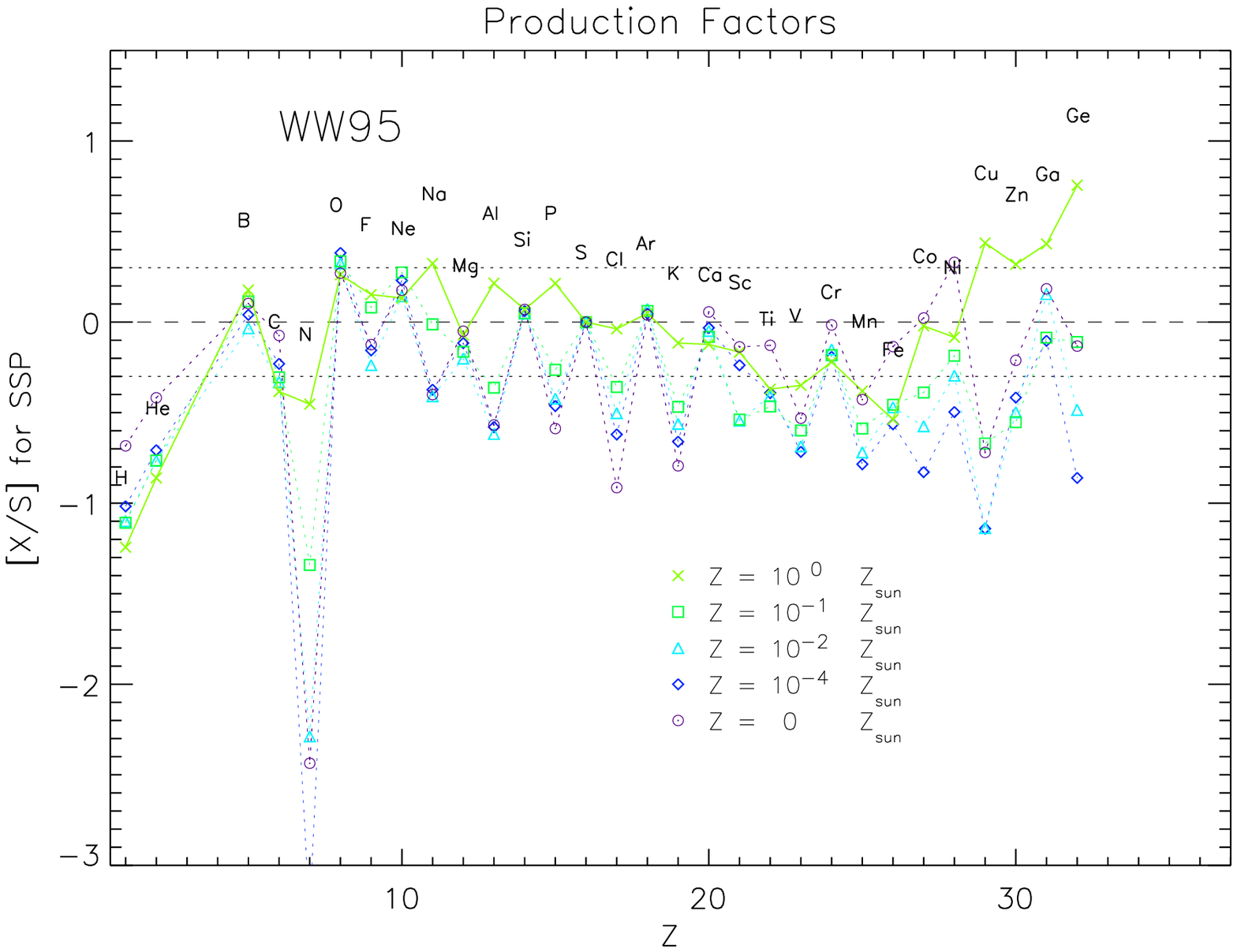}
\caption{Production factors relative to S from a
  single generation of massive stars from 11--100\,{\msun} on a solar
  logarithmic scale using the yields of Woosley \& Weaver 1995.  Five
  metallicities are shown from zero to solar. C, N and some of the
  iron-peak elements are subsolar because they require additional
  sources such as intermediate- and low-mass stars and Type~Ia SNe.
   \label{PFfig}}
\end{figure*}

The model results displayed in Figure~\ref{Zfig} reveal the so-called
``odd-even'' effect, whereby the underabundance of odd-numbered
elements with respect to their even-numbered neighbours increases with
decreasing metallicity. This effect is a signature of massive SNe from
subsolar-metallicity stars. Figure~\ref{PFfig} shows the production
factors relative to S from the WW95 SN~II models integrated over the
Kroupa et~al. (1993) IMF over the mass range 11-100\,M$_{\odot}$. The
odd-even effect persists from the zero to 1/10th solar metallicity
WW95 models, with the trend reversing in the solar model. The strength
of the odd-even effect in this DLA can be gauged by comparing
abundances between pairs of neighbouring odd-Z and even-Z elements.
Evidence for a mild odd-even effect comes from the ratios
[P/Si]~$<$~$-$0.15, [Mn/Fe]~$<$~$-$0.16, and [Ga/Ge]~$<$~$-$0.13.

Even-elements behave as primary products in SNe~II and their stellar
yields are robust to the initial metallicity.  Conversely, the ratios
of odd-elements to S show spreads of typically $\sim$~0.2~dex over the
0.4~dex range in [O/H]. The SNe~II yield of odd-elements from fluorine
to the iron-peak is understood to be sensitive to the neutron excess
and therefore increases with the initial abundance of ``seed'' nuclei
present in the star. Thus the ratio of odd-elements to S is highest in
the [O/H]~=~$-0.24$ model.

Copper exhibits the greatest sensitivity to metallicity, with Cu/S
having a stronger than linear dependence on O/H over the metallicity
range considered. Thus [Cu/S] has the potential to indirectly
constrain the metallicity and the H\,\textsc{i} column density. The
upper limit on the Cu detection arises because of line blending.
However the depletion of Cu onto dust grains reduces its abundance in
the gas-phase. Accounting for these opposing effects, the intrinsic Cu
abundance is likely to fall within 0.4~dex of the current upper limit.
Only the highest metallicity models approach the Cu observation.  The
predicted [Zn/S] value, which is underproduced by $\sim$0.4~dex in the
[O/H]~=~$-$0.44 model, is also in closer agreement with data for the
highest metallicity model.

Abundances of the odd-elements Al, P, and Cl relative to S offer a
probe into the nature of the ``odd-even'' effect. However, only lower
(for Al and Cl) and upper limits (for P) exist at present in this DLA.
These limits are currently not sufficiently restrictive to eliminate
any of the three different metallicity models depicted in
Figure~\ref{Zfig}. Owing to the strength of the
Al~\small{II}~$\lambda$~1670 line, only lower limits can be placed on
the Al abundance in many DLAs, yet absorption-line systems present a
unique opportunity to acquire empirical evidence for the
metallicity-dependent behaviour of P and Cl, whose weak stellar
spectral lines mean there is a paucity of data from local stars. A
database of measurements of these elements in DLAs will help test the
theoretical expectation that Cl, P, and Al follow similar evolutionary
paths.

One might expect from the theoretical predictions shown in
Figure~\ref{Zfig} that improved observations of P and Cl will see both
[P/S] and [Cl/S] converging toward $\sim -0.4$. However, the abundance
of Cl shown in these figures was derived from the Cl~\small{I}
transition, which is not the dominant ion. After correcting for
ionisation, the true Cl abundance is likely more than 0.5~dex higher
than the current lower limit.  Even when these models are evolved to
solar metallicity, Al and Cl are still underproduced by 0.1 and
0.3~dex, respectively; suggesting that the odd-even effect may be
slightly exaggerated in the stellar models, or that intermediate-mass
stars provide an additional source. It should be noted that the solar
reference Cl abundance used in this paper is $\sim 0.15$~dex higher
than the Anders \& Grevesse (1989) value.

Boron is synthesised in a variety of ways including neutrino-induced
reactions in the shells of SNe~II (the $\tau$-process) and cosmic ray
spallation onto seed nuclei in the interstellar medium (Fields \&
Olives 1999). The relative importance of these processes is expected
to vary with metallicity. The $\tau$-process should produce constant
B/O regardless of metallicity, whereas a signature of the cosmic ray
process would be B/O decreasing with decreasing metallicity. The WW95
models generate B via the $\tau$-process and the satisfactory
agreement with the observed B/O ratio lends support to a
$\tau$-process origin for B.

\subsubsection{Sensitivity to the Stellar Initial Mass Function}\label{sect_imf}

\begin{figure*}
%\plotone{ChemComp_WW_vs_mU_wrtS.eps}
%\epsscale{1.24}
\plotone{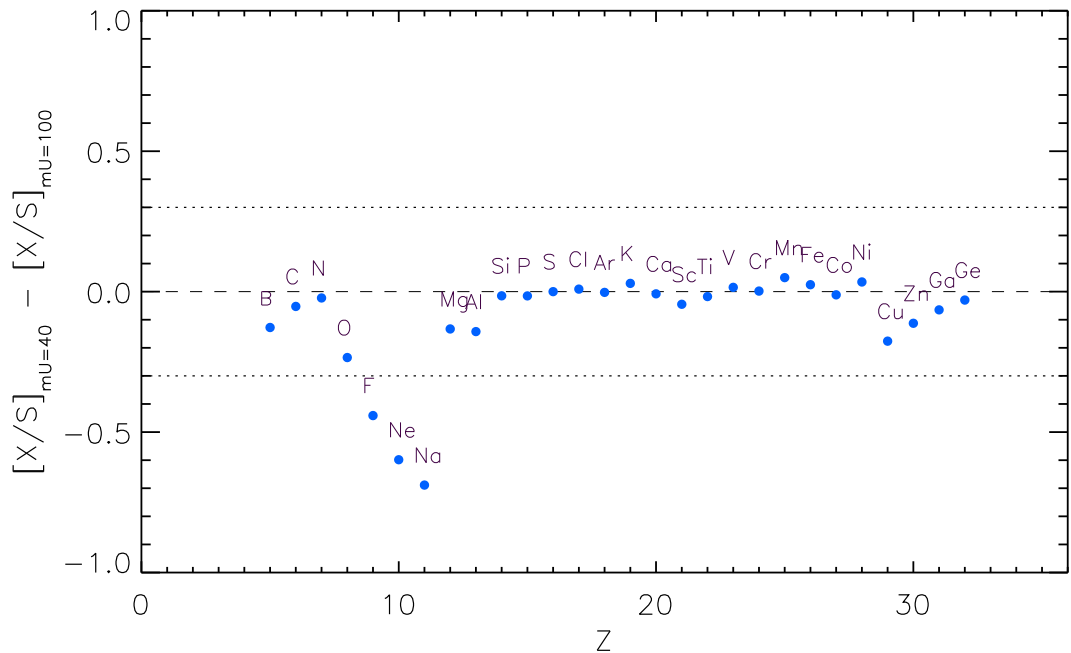}
\caption{Logarithmic ratio of the abundance pattern obtained with an
  IMF upper limit of 40\,M$_{\odot}$ and with an upper limit of
  100\,M$_{\odot}$.
    \label{IMFfig}}
\end{figure*}

It has been estimated that uncertainties in the shape and limits of
the IMF correspond to variations in the \emph{absolute} elemental
yields by a factor of 2 (Wang \& Silk 1993).  We have tested the
sensitivity of our results to a change in the upper mass limit of the
IMF from 100 to 40\,{\msun}. Figure~\ref{IMFfig} plots the logarithmic
ratio of the abundance pattern obtained in these two cases. The
abundances of most elements relative to S are unchanged. Prominent
exceptions are F, Ne, and Na, whose yields relative to S decrease by
factors ranging from 3\,-\,5. [O/S] is also lower by $\sim$0.25~dex
when the IMF is curtailed at 40\,M$_{\odot}$. The magnitude of the
offset reflects the dependence of elemental yield on stellar mass. The
elements most affected by the shift in upper IMF limit are those whose
yields increase most steeply with initial stellar mass. The possible
future detection of F in this DLA may help constrain the IMF.

\subsubsection{Sensitivity to the Type II Supernova Nucleosynthesis Prescriptions}\label{sect_wwvsfr}

\begin{figure*}
%\plotone{ChemComp_wrtS_WWvsFRANEC.eps}
%\epsscale{1.25}
\plotone{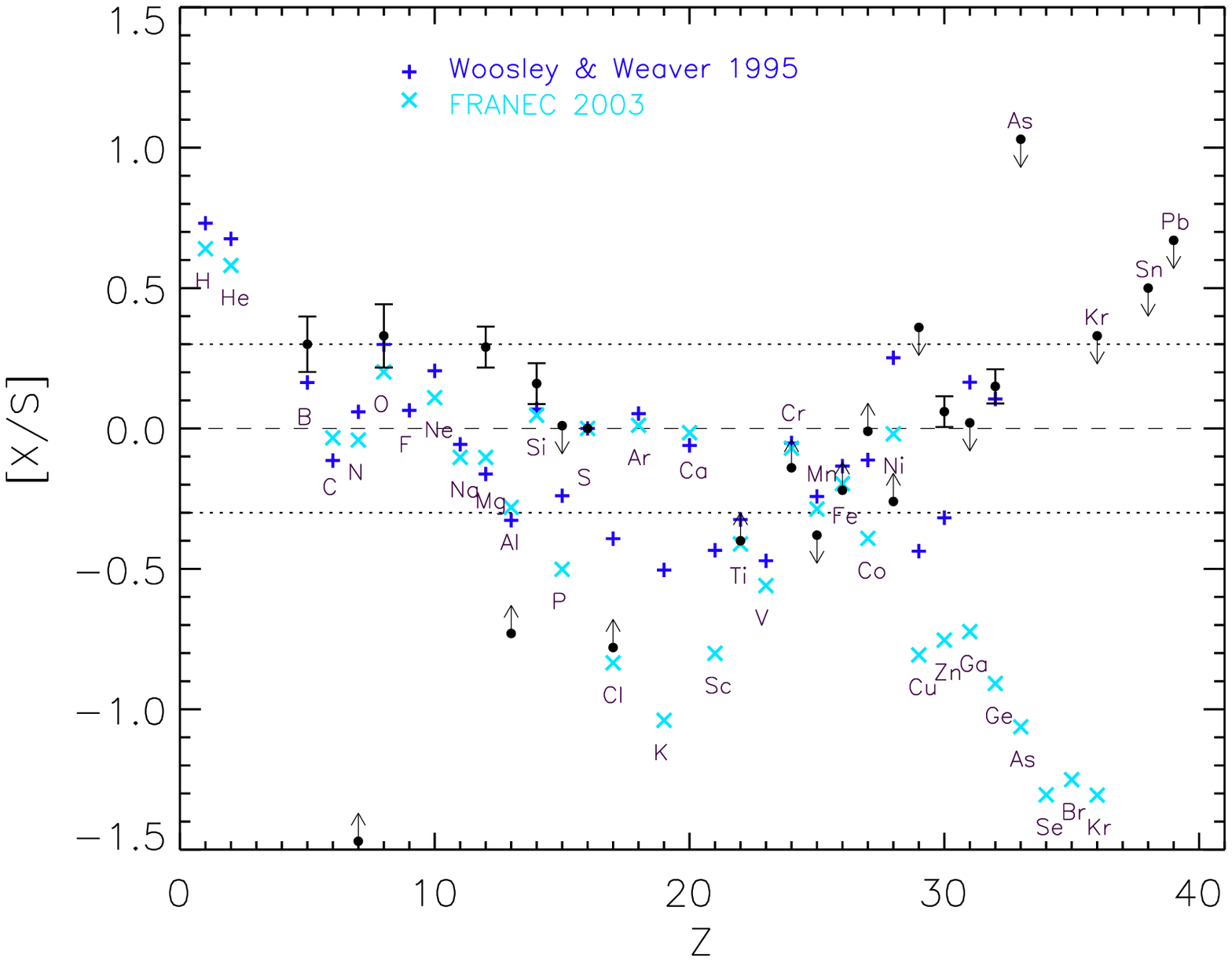}
\caption{Predicted abundance pattern of a 2.2~Gyr
  object with [O/H]~=~$-$0.44 obtained using the Woosley \& Weaver
  1995 yields (\emph{plus signs}) and FRANEC 2003 yields
  (\emph{crosses}).  Other symbols have the same meaning as in
  Figure~\ref{agefig}.
    \label{WWvsFRfig}}
\end{figure*}

Figure~\ref{WWvsFRfig} compares the standard WW95 model with the same
model using FRANEC yields for massive stars. We note several main
differences: 1) B and F are produced in negligible amounts in the
FRANEC model because neutrino-induced nucleosynthesis processes are
omitted from their calculations; 2) the elements from Cu to Kr are
present in trace amounts in the FRANEC model; and 3) the odd-even
effect is much more pronounced in FRANEC models. In particular, P, Cl,
K, and Sc are significantly lower with respect to S in the FRANEC
model, whereas the abundance of their even-Z neighbours matches those
from the WW95 model. The deficit of Cl from the FRANEC models with
respect to the observations suggests that the odd-even effect may be
too severe. Despite these differences, there is good consensus between
the two models for about half of the elements with DLA measurements
and many of the conclusions of this paper can be reached irrespective
of the choice of SN~II yields.

\subsection{Discussion of Individual Elements}

\subsubsection{Oxygen}

In the most widely cited set of solar abundances from Anders \&
Grevesse (1989), the solar oxygen abundance was assumed to be
log(N$_{\mathrm{O}}$/N$_{\mathrm{H}}$)$_{\odot}$ + 12 = 8.93. However,
recent estimates that account for solar granulation and non-LTE
effects lead to significant downward revisions in the Sun's oxygen
abundance by almost a factor of two (to
log(N$_{\mathrm{O}}$/N$_{\mathrm{H}}$)$_{\odot}$ + 12 = 8.69 - Allende
Prieto, Lambert \& Asplund 2001). Such a shift partly resolves the
long-standing dichotomy between the Sun's oxygen abundance and that of
the local ISM (Andr{\' e} et al. 2003). This work adopts Holweger's
(2001) preferred value of
log(N$_{\mathrm{O}}$/N$_{\mathrm{H}}$)$_{\odot}$ + 12 = 8.73, whereas
WW95 and FRANEC stellar yields were derived with the Anders \&
Grevesse (1989) values. Owing to uncertainties in stellar
nucleosynthesis inputs such as the $^{12}$C($\alpha,\gamma$)$^{16}$O
reaction rate, convection, and mass-loss (see Langer 1996; Gibson,
Loewenstein, \& Mushotzky 1997 for useful discussions of these
uncertainties), we consider our predictions for the O abundance to be
accurate within a factor of $\sim$~2.

\subsubsection{Magnesium}\label{mg_sect}

Figure~\ref{modelvsDLAfig} shows that the [Mg/S] value predicted by
the standard model is 0.45 dex lower than is observed in this DLA.
The reason is clear from Figure~\ref{PFfig}, which shows that Mg/S is
slightly subsolar in the ejecta from a generation of WW95 metal-poor
stars and tends toward the solar value at higher metallicity. The
observed value of [Mg/S]$_{\mathrm{DLA}}$~=~+0.3~dex cannot be matched
by any single generation of WW95 stars and nor will any superposition
of populations fit the data. Alib{\' e}s et~al. (2001) also found that
WW95 yields lead to an underproduction of Mg in their models of the
local disk; a problem which they suggest could be resolved if
intermediate-mass stars or SNe~Ia supply additional Mg. Indeed,
4~$-$~6~{\msun} stars have been shown to contribute to most of the
abundance of the neutron-rich Mg isotopes at low and intermediate
metallicities (Fenner et al. 2003).

\subsubsection{Sulfur}

Sulfur was used as the reference element in this study in preference
to Fe because its nucleosynthetic origin is thought to be better
understood. Hydrostatic burning and explosive O and Si burning in
massive stars are the main processes responsible for S production.
Although Fe has traditionally been a popular metallicity gauge, owing
to easily observable stellar spectral lines, the question of its
origin is complicated by the fact that both Type~Ia and Type~II SNe
can make important contributions to the interstellar Fe content.
Moreover, the Fe yield from massive stars is subject to uncertainties
relating to the SN~II mass cut. Oxygen is a common alternative
reference element, but due to the recent significant revisions in the
solar O abundance, we chose to plot abundance ratios relative to S,
which has been detected in a reasonable number of DLAs and local
stars.

\subsubsection{Titanium}

Titanium is a curious element given that observationally, it follows
the trends of the alpha-elements in nearby stars, but theoretically,
is expected to behave as an iron-peak element. There is currently only
a lower limit on Ti in this protogalaxy, however one might expect Ti
to be enhanced with respect to iron, given the enhancement of O, Mg,
and Si. In contrast, the model predicts [Ti/Fe]~$\sim$~-0.2~dex. The
deficit of Ti generated by standard SN~II and SN~Ia models is a
well-known unresolved problem (e.g.  Timmes et~al.  1995, Alibes
et~al.  2001).

\subsubsection{Zinc} \label{zn_sect}

Iron is often used as a metallicity gauge in stellar populations
because of the ease with which it can be observed. However, the
abundance of iron in gas is difficult to estimate because it is
readily incorporated into dust grains (Savage \& Sembach 1996).
Furthermore, all of the elements comprising the Fe-peak are
refractory.  Thus Zn has become a popular substitute for Fe in DLA
studies because it is largely free from the effects of dust-depletion
and because the Zn/Fe ratio in local stars is approximately solar over
a wide range of metallicities.  The constancy of Zn/Fe vs Fe/H has
lead authors to propose that Zn, like Fe, is generated mostly by
Type~Ia SNe (e.g.  Matteucci et al.  1993, Mishenina et~al. 2002).
However this need not be the case, since the metallicity dependence of
Zn production in massive stars can naturally lead to a time delay that
mimics that of SN~Ia. As can be inferred from Figure~\ref{agefig}, Zn
yields predicted by the W7 Type~Ia SNe model (Iwamoto 1999) are small
in comparison to WW95 SNe~II yields (whereas the iron-peak elements
from V to Ni can be largely attributed to SNe~Ia).  Figure~\ref{Zfig}
reveals the strength of the metallicity-dependence of Zn production in
massive stars. These effects can combine to allow Zn and Fe enrichment
to occur in lockstep despite having different nucleosynthetic origins.

\begin{figure*}
%\plotone{SFeZn_behaviour.eps}
%\epsscale{0.9}
\plotone{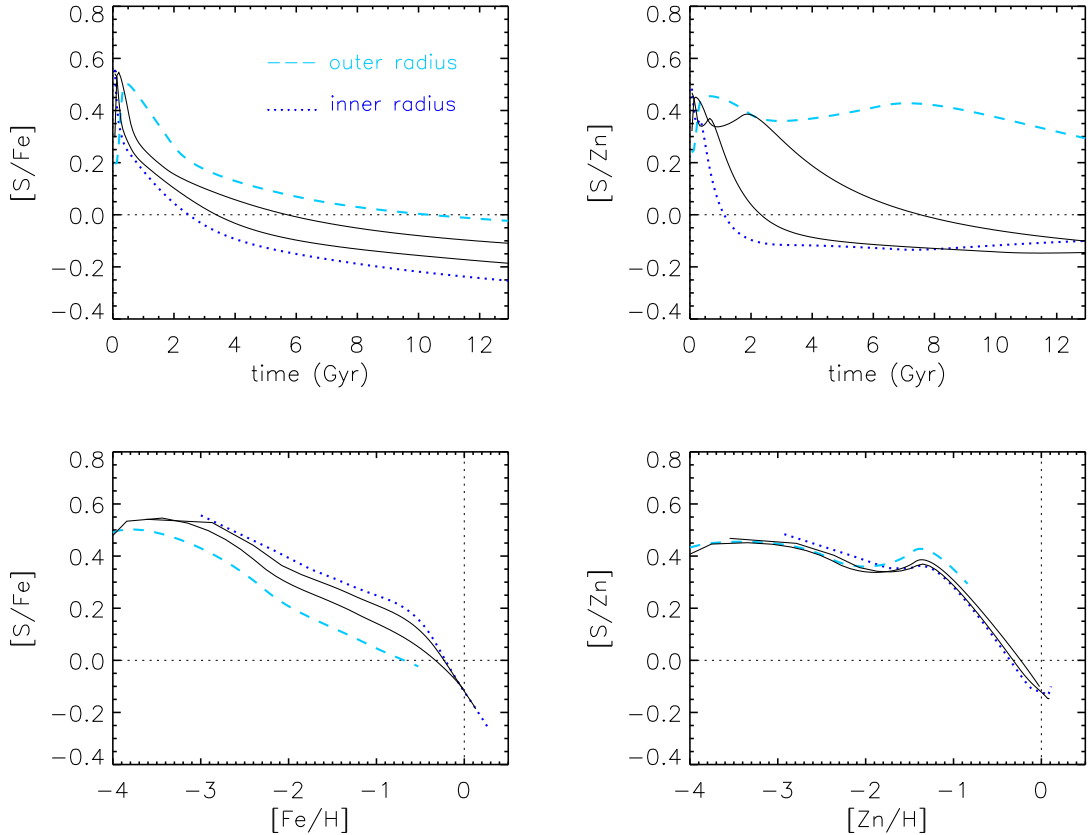}
\caption{The evolution of S relative to Fe (\emph{left panels}) and relative
  to Zn (\emph{right panels}) as a function of time (\emph{upper
    panels}) and metallicity (\emph{lower panels}). The four curves in
  each figure correspond to different radii in a Milky Way-like disk
  galaxy. The inner radius (\emph{dotted line}) reaches subsolar
  [S/Fe] quickly owing to earlier and stronger star formation. The
  outer region of the disk (\emph{dashed line}) is expected to remain
  enhanced in S and alpha-elements for longer because the drawn-out
  star formation causes the SNe~Ia rate to peak at later times. In
  [Fe/H]-space the trends are reversed, with the inner disk
  maintaining supersolar [S/Fe] at higher metallicity because the
  early intense star formation pushes the gas towards high [Fe/H]
  before SNe~Ia start to dominate Fe production. The upper panels show
  that only for certain radii does the evolution of Zn resemble that
  of Fe. Given our adopted SN~II and SN~Ia nucleosynthesis
  prescriptions, Zn originates primarily from SNe~II with a
  metallicity-dependent yield. Sulfur also comes mostly from SNe~II
  with a fairly metallicity-independent yield. Thus the curves showing
  [S/Zn] vs [Zn/H] for various radii overlay one another.
   \label{SFeZnfig}}
\end{figure*}

Figure~\ref{SFeZnfig} illustrates potential problems associated with
using Zn as a proxy for Fe. The four panels plot [S/Fe] (\emph{left
panels}) and [S/Zn] (\emph{right panels}) versus time (\emph{upper
panels}) and metallicity (\emph{lower panels}). The curves correspond
to different radii in a Milky Way-like galaxy. The inner radius has
the strongest star formation rate and the earliest peak.  Star
formation proceeds more slowly with increasing radius. Owing to the
delayed release of large amounts of Fe from SNe~Ia with respect to the
release of S from short-lived massive stars, the behaviour of [S/Fe]
(or [O/Fe]) vs [Fe/H] can be used to diagnose star formation
histories. Slower and more protracted star formation, such as in the
outer disk of a galaxy, should to lead to subsolar [S/Fe] (and [O/Fe])
at lower [Fe/H] (\emph{lower left panel}). But when Fe is replaced by
Zn (\emph{lower right panel}), the behaviour at all radii is virtually
indistinguishable. This is a consequence of the nucleosynthesis
prescriptions adopted in this paper that have Zn being produced mostly
by massive stars as an increasing function of initial stellar
metallicity. Under this formalism, the evolution of Zn mirrors that of
Fe only for certain star formation histories, but might deviate
markedly in other environments. Note that it is impossible to attain
the solar and subsolar values of S/Zn at the low Zn/H that are
observed in DLAs (e.g. Centuri{\' o}n et al. 2000) using the
conventional Type~Ia and II SNe models employed in this study,
regardless of SF history (\emph{lower right panel}). This may indicate
a deficiency in our theories of Zn production from the standard SN~II
and SN~Ia nucleosynthesis models. Other authors (e.g. Calura,
Matteucci, \& Vladilo 2003) have produced low [S/Zn] at low
metallicity using chemical evolution models, but they achieve this by
ignoring the theoretically predicted SN~II and SN~Ia Zn yields and
instead assume that Zn comes mostly from SNIa and scales with the Fe
yield.

There are two main weaknesses of the standard models of Type~Ia and II
SNe that highlight the uncertainties in the nucleosynthetic origins of
Zn: firstly, the models are unable to explain the supersolar values of
[Zn/Fe] in metal-poor stars; and secondly, the predicted Zn isotopic
composition is at odds with solar. Breaking down the theoretical
SNe~II Zn yield into its main isotopes reveals substantial
disagreement with the solar pattern. The dominant isotope in the Sun
is $^{64}$Zn, yet WW95 Type~II models underestimate its abundance by a
factor of $\sim$3. Conversely, $^{68}$Zn is overproduced by a similar
factor. Since the W7 SNe~Ia model also underpredicts $^{64}$Zn, other
processes might need to be invoked to elevate $^{64}$Zn to solar
proportions. The uncertainty regarding zinc's production sites is
unfortunate given its key role in diagnosing DLAs.

Neutrino-driven neutron star winds (Woosley \& Hoffman 1992; Hoffman
et~el. 1996) may be an important supplemental site for Zn synthesis.
WW95 claim that neutrino winds accompanying the r-process are probably
the chief production site of $^{64}$Zn, as opposed to the neutron
capture process during helium burning that is included in their SN~II
models. Alternatively, a blend of fallback and mixing in energetic
hypernovae (HNe) have been found to produce $^{64}$Zn in prodigious
amounts (Umeda \& Nomoto 2002), leading to [Zn/Fe] values compatible
with stellar observations. Each of these HNe may eject up to
4$\times10^{-4}$ M$_{\odot}$ of $^{64}$Zn; about ten times more than a
typical SN~II.

Figure~\ref{SFeZnfig} illustrates the tendency for a system with a
``burstier'' SFH (such as the innermost disk) to maintain elevated
[S/Fe] over a wide range of [Fe/H] because the ISM becomes metal-rich
before SNe~Ia have a chance to dominate Fe production. The
age-metallicity relation is gradual in the outer disk, such that for a
given [Fe/H], there has been more time for SNe~Ia to contribute to the
abundance pattern. Thus, the outer radii in Figure~\ref{SFeZnfig}
reaches [S/Fe]~=~0 at [Fe/H]~=~$-$0.7, whereas the inner radii has
[S/Fe]~=~+0.2 at [Fe/H]~=~$-$0.7. Note however, that in the time
domain, it is the slowly forming systems that maintain elevated [S/Fe]
for longer temporal periods. Indeed, it takes about 10~Gyr for the
outer disk in this model to reach solar S/Fe. This corresponds to a
much lower redshift than is characteristic of most DLAs. The DLA
population is characterised by roughly solar $\alpha$/Fe values, a
mean metallicity $\sim$~1/10th solar, and ages $\lesssim$~5~Gyr.
According to our standard disk galaxy model, in which the gas infall
timescale increases with Galactocentric radius, only the outer regions
of the disk can reach [$\alpha$/Fe]~$\sim$~0 at [Fe/H]~$\sim$~$-$1,
but on timescales $>$~5~Gyr. In order to get [$\alpha$/Fe]~$\sim$~0 at
[Fe/H]~$\sim$~$-$1 on sufficiently short timescales one may invoke a
short star formation burst of low efficiency (to keep the metallicity
down) or a galactic wind that preferentially removes $\alpha$-elements
(e.g.  Calura et~al. 2003). Dwarf galaxies are more closely associated
with these phenomena than disk galaxies.

\subsubsection{Gallium and Germanium} 

The predictions from our standard model for the abundance of Ga and Ge
are in rough agreement with the data, however we hesitate to draw
conclusions based on these results owing to the uncertainties
surrounding their synthesis. Although Woosley \& Weaver (1995) include
Ga and Ge in their tables of yields, they state that the synthesis of
all isotopes above $^{66}$Zn are not considered accurate. Indeed, the
WW95 Ga and Ge yields are remarkably sensitive to factors including
the explosion energy.  We also note that WW95 predict a
\emph{stronger} metallicity-dependence for the even-numbered Ge than
for the odd-numbered Ga (Figure~\ref{PFfig}).

\section{Summary and Future Directions}

We have predicted the chemical evolution of a DLA as a function of
star formation history, age, and metallicity, and find that the
enrichment pattern detected in absorption in a $z$\,=\,2.626
protogalaxy is generally consistent with the nucleosynthetic signature
of Type~II SNe with a moderate contribution from Type~Ia SNe. Despite
this gratifying agreement with the data, a few inconsistencies remain
that may provide insight into nucleosynthetic processes. The
underproduction of Mg and Zn with respect to S in our models lends
support to an idea already hinted at by discrepancies between the
observed abundances in nearby stars and the yields from standard SN~II
models, that these elements may require additional production sites
such as: intermediate mass stars, in the case of Mg; and
neutrino-winds, in the case of Zn. It is also likely that tighter
constraints on Cl and Cu in the future will see these elements being
underproduced by all but the highest metallicity SN~II models. This
may indicate that the odd-even effect is weaker than predicted by
stellar nucleosynthesis models.

{\dlan} is expected to soon be complemented by a growing database of
detailed abundance patterns of DLAs covering a range of metallicities
and redshifts. These observations will represent a new regime for
probing chemical evolution in diverse environments with an assortment
of enrichment histories. As well as providing further insight into the
nature of the DLA population, such a database will complement local
stellar observations to help uniquely constrain nucleosynthesis
processes.  In particular, we await future detections of Cl and P in
this and other DLAs. Detection of these odd-elements should enable the
SN~II models of WW95 and FRANEC to be assessed in terms of their
different predictions for the magnitude of the odd-even effect. We
also propose that the detection of s-process elements like Kr and
particularly Pb, holds promise as a sensitive measure of the age of
intermediate to high redshift DLAs.

Future observations of {\dlan} are also expected to yield measurements
of C, N, F, Ga, and Sn.  Fluorine, like boron, is difficult to measure
in stars, but can be detected in absorption in DLAs, providing a test
of the $\tau$-process believed to operate during the core-collapse
phase of massive stars. It is hoped that Ga measurements will provide
insight into the r-process in the early universe. Carbon and nitrogen
detections will be especially fruitful because they can be compared
against the dataset of N in DLAs and C and N in local stars.
Furthermore, knowledge of CNO abundances in a single DLA will
constrain theories of nucleosynthesis in intermediate mass stars.

\acknowledgments

The authors would like to thank A. Wolfe and J.C. Howk for allowing us
to present results from the {\dlan} observations prior to full
publication. We are grateful to Alessandro Chieffi and Marco Limongi
for providing their latest stellar yields prior to publication. We
wish to thank Stan Woosley for helpful discussions.  BKG acknowledges
the support of the Australian Research Council, through its Large
Research Grant and Discovery Project schemes. We thank the referee
P. Bonifacio for his detailed comments that helped improve this paper.

\end{document}